\documentclass[journal,twoside,web]{ieeecolor}
\usepackage{mathrsfs}  

\usepackage{generic}
\usepackage{epsfig} 
\usepackage{times} 
\usepackage{amsmath} 
\usepackage{amssymb}  
\usepackage{algorithmic}
\usepackage{subfigure}
\usepackage{color}

\usepackage{algorithm}
\usepackage{bm}
\usepackage{graphicx}
\usepackage{multirow}
\usepackage{array}
\usepackage{graphics} 

\allowdisplaybreaks
\floatname{algorithm}{\emph{Algorithm}}
\renewcommand{\algorithmicensure}{Output:}
\renewcommand{\algorithmicrequire}{Input:}

\newcommand{\cS}{\mathcal{S}}

\newtheorem{theorem}{Theorem}
\newtheorem{proposition}{Proposition}

\newtheorem{definition}{Definition}

\newtheorem{example}{Example}

\title{CA-Observability of Discrete Event Systems under Cyber Attacks}

\author{Shengbao Zheng, Shaolong Shu, \IEEEmembership{Senior Member, IEEE} and Feng Lin, \IEEEmembership{Fellow, IEEE}
		\thanks{The authors are supported by the National Natural Science Foundation of China under Grants 62073242 and by the National Science Foundation of USA under grant 2146615.}
		\thanks{Shengbao Zheng is with the School of Automation and Electrical Engineering , Shenyang Ligong University, Shenyang 110159, Liaoning, China(E-mail: Zheng\_shengbao@sylu.edu.cn).}
		\thanks{Shaolong Shu is with the School of Electronics and Information Engineering, Tongji University, Shanghai 200092, China (E-mail: shushaolong@tongji.edu.cn).}
		\thanks{Feng Lin is with the Department of Electrical and Computer Engineering, Wayne State University, Detroit, MI 48202 USA (E-mail: flin@wayne.edu).}}
\def\BibTeX{{\rm B\kern-.05em{\sc i\kern-.025em b}\kern-.08em
		T\kern-.1667em\lower.7ex\hbox{E}\kern-.125emX}}
\begin{document}

\maketitle

\begin{abstract}	
In a previous paper, we investigate the control problem of discrete event systems under cyber attacks, where CA-controllability and CA-observability are proposed. 
We prove that a discrete event system can be controlled to generate a specification language $K$ if and only if $K$ is CA-controllable and CA-observable. 
While CA-controllability can be ``converted'' to (conventional) controllability, CA-observability cannot be ``converted'' to (conventional) observability. In this paper, we further investigate CA-observability. 
We develop a method and an algorithm to check CA-observability. The method involves constructing an augmented automaton whose states are pairs of the current state and state estimate of the supervised system. 
We prove that $K$ is CA-observable if and only if the language generated by the augmented automaton is equal to $K$. An algorithm is then developed to check CA-observability. 
We also investigate the properties of CA-observability. We show that CA-observability is not closed under (set) intersection. Note that (conventional) observability is closed under intersection. 
We further show that CA-observability is not closed under union. This is similar to observability. 
If $K$ is not CA-controllable and CA-observable, we propose a method to calculate a CA-controllable and CA-observable sublanguage. We first calculate the supremal CA-controllable sublanguage for  $K$ which can be described by a sub-automaton of the automaton for $K$. We then propose a state-estimate-based supervisor based on the sub-automaton, which ensures the supervised system stays within $K$.
\end{abstract}

\section{Introduction} \label{s1}

Many practical engineering systems, such as power systems, manufacturing systems, computer systems, transportation systems, and smart homes can be modeled as discrete event systems (DES) at some level of abstraction. 
The main control theory developed for DES is the supervisory control theory \cite{ramadge1987supervisory, lin1988observability, rudie1992think, cassandras2021introduction, wonham2019supervisory}. 
The cornerstones of supervisory control are controllability and observability. 
Together, they characterize the necessary and sufficient condition for the existence of a controller, called supervisor, such that the closed-loop system satisfies control requirement specified by a specification language.

Because of the importance of controllability and observability, many works have been done to investigate the properties of controllability and observability. 
It is shown that controllability and observability can be checked with polynomial complexity \cite{ramadge1987supervisory}, \cite{rudie1995computational}. 
It is also shown that controllability is closed under set union and hence the supremal controllable sublanguage of any specification language exists and is unique \cite{ramadge1987supervisory}. 
It is further shown that observability is closed under set intersection and hence the infimal observable superlanguage of any specification language exists and is unique \cite{lin1988observability}. 
Based on these results, an ``optimal'' supervisor can be synthesized such that the closed-loop system satisfies legal requirement and the supervisor gives the system maximum freedom to execute events. 
This maximum freedom often leads to best performance of the system \cite{lin1994linguistic}.

Since the invention of the Internet, networks have become  inseparable parts of engineering systems, including control systems. 
Nowadays, many controllers are connected to systems they control (plants) by shared wired or wireless communication networks. 
Compared with non-networked control, networked control faces at least two more challenges: 
(1) In networked control, communication delays and losses between controllers and plants are unavoidable. Since communication delays and losses have significant impacts on control, how to design controllers that are robust with respect to communication delays and losses is a big challenge. 
(2) Because communication networks are often shared by many users in networked control, cyber attacks are much more likely to occur in networked control. How to design controllers that are robust against cyber attacks is another big challenge.

Both challenges have been investigated in control of networked discrete event systems. For control of networked DES with communication delays and losses, we refer the reader to the following survey and tutorial papers. 

Survey paper \cite{zhu2023overview} presents an overview of the networked supervisory control for network DES for both centralized supervisory control and decentralized supervisory control. 
The paper surveys both observation delays and control delays, both FIFO channels and non-FIFO channels. Communication losses are also surveyed.

Survey paper \cite{liu2020modeling} gives a comprehensive survey on modeling, analysis and control of networked DES. 
For networked models, \cite{liu2020modeling} classifies the models in existing literature by three criteria. 
For analysis of network properties, \cite{liu2020modeling} surveys network observability, network controllability, and other network properties. 
For supervisory control, \cite{liu2020modeling} classifies the control problems into supervisor existence problem, and supervisor synthesis problem.

The tutorial paper \cite{lin1999modeling} gives a comprehensive coverage of an approach to modeling and control of networked DES proposed by us. It includes modeling of networked DES, nondeterministic supervisory control, deterministic supervisory control, nonblocking supervisory control, robust supervisory control, and decentralized supervisory control. 

For control of networked discrete event system under cyber attacks, extensive works have been done, for example, in \cite{fritz2018modeling, wakaiki2019supervisory, zhang2018stealthy, wang2019supervisory, khoumsi2019sensor, carvalho2016detection, basilio2021analysis}.

A good overview of modeling and control of networked DES under cyber attacks can be found in \cite{rashidinejad2019supervisory}, where the authors classify the works under a framework consisting of three dimensions: (1) the communication channel where an attack can happen, (2) the attack impact on the transmitted data, and (3) the mechanism to prevent damage. 

Following the publication of \cite{rashidinejad2019supervisory}, subsequent studies have expanded control of networked discrete event system under cyber attack. In \cite{wang2020mitigation}, a framework employing probabilistic discrete event systems incorporates a probability measure for unsafe behavior into attack models. The focus is on addressing the least-unsafe supervisor problem to minimize this measure, thereby enhancing existing attack mitigation techniques.

In the realm of cyber attacks on observation channels,  \cite{alves2022discrete} delves into scenarios where attackers can insert or erase events. Two algorithms and a novel approach are presented to assess the feasibility of a supervisor enforcing a specified legal language characterized by P-observability.

The work in \cite{zhang2021joint} centers on state estimation in networked DES cyber attacks. Utilizing a joint estimator with two state observers, an attacker observer and an operator observer, the authors aim to derive sets of states consistent with both uncorrupted and corrupted observations.


We also proposed an approach in \cite{zheng2021modeling, zheng2023modeling} to investigate modeling and control of discrete event systems under cyber attacks.
We use a model called ALTER (for Transition-basEd Replacement using Attack Languages) to describe cyber attacks on sensors, where an attacker can replace an observed transition by one of the strings in a given attack language.
In this way, sensor attacks are modeled by a set of attack languages, which are specified by a model builder based on the knowledge of the attacker. In particular, deletions, insertions, and replacements can all be modeled by some attack language.
For cyber attacks on actuators, we assume that an attacker can alter control commands issued by a supervisor by changing the disablements and enablements of attacked events.

We show in \cite{zheng2023modeling} that, due to nondeterminism in sensor attacks and actuator attacks, the language generated by the supervised system is not unique.
To ensure the safety of the supervised system, an upper bound on the set of languages generated by the supervised system, called large language, is defined and investigated.
We extend controllability and observability to CA-controllability and CA-observability. We then show that the existence conditions for a robust supervisor whose large language is equal to a specification language $K$ is that $K$ is CA-controllable and CA-observable.

It turns out that while CA-controllability can be ``converted'' to (conventional) controllability, CA-observability cannot be ``converted'' to (conventional) observability. Hence, the methods developed for observability cannot be extended for CA-observability. 

In this paper, we propose to further investigate CA-observability. We develop a method and an algorithm to check CA-observability. The method involves constructing an augmented automaton whose states are pairs of the current state and state estimate of the supervised system. We prove that this augmented automaton generates the large language. We further prove that $K$ is CA-observable if and only if the language generated by the augmented automaton is equal to $K$. An algorithm is then developed to check CA-observability. 

We also investigate the properties of CA-observability. We show that CA-observability is not closed under (set) intersection. Note that (conventional) observability is closed under intersection. Therefore, while the infimal observable superlanguage of $K$ always exist, the infimal CA-observable superlanguage of $K$ may not exist. We also show that CA-observability is not closed under union. This is similar to observability. Therefore, the supremal CA-observable sublanguage of $K$ may not exist. 

If $K$ is not CA-controllable and CA-observable, then in order to design a robust supervisor, we  need to find a CA-controllable and CA-observable sublanguage of $K$. To this end, we first calculate the supremal CA-controllable sublanguage $K^{\uparrow}$ of $K$ which can be described by a sub-automaton $H^{\uparrow}$ of the automaton for $K$. We then propose a state-estimate-based supervisor based on $K^{\uparrow}$. We prove that the large language of the controlled system is a suitable CA-controllable and CA-observable sublanguage. 

Compared with \cite{zheng2023modeling}, the paper is novel in the following aspects. (1) An approach for checking CA-observability is proposed for the first time. (2) We prove that CA-observability is not closed neither under intersection nor under union. (3) We propose a novel approach to calculate a CA-controllable and CA-observable sublanguage of a given $K$. In this way, we also obtain a state-estimate-based supervisor to control the given system to generate such a sublanguage.   

The paper is organized as follows. In Section II, we briefly review the results of \cite{zheng2023modeling} with some improvements. In Section III, we present the method and algorithm for checking CA-observability. In Section IV, we investigate whether CA-observability is closed under intersection and/or union. In Section V, we propose a novel approach to calculate a CA-controllable and CA-observable sublanguage and a state-estimate-based supervisor to generate such a sublanguage. We conclude the paper in Section VI. 

\section{Supervisory Control of Discrete event systems under cyber attacks} \label{s2}

\subsection{Discrete Event Systems}

In this section, we review supervisory control of discrete event systems under cyber attacks \cite{zheng2021modeling, zheng2023modeling}.
A discrete event system is modeled as a deterministic automaton \cite{cassandras2021introduction}
$$
G=( Q, \Sigma , \delta , q_0, Q_m ),
$$
where $Q$ is the set of states; $\Sigma$ is the set of events; $\delta : Q \times \Sigma \rightarrow Q$ is the (partial) transition function; $q_0$ is the initial state; and $Q_m$ is the marked state set. 
In this paper, we assume that all states in $G$ are marked, that is, $Q_m=Q$.
The set of all possible transitions is also denoted by $\delta$, that is, $\delta = \{ (q, \sigma, q'): \delta (q, \sigma) =q' \}$.

We use $\Sigma ^*$ to denote the set of all strings over $\Sigma$. The transition function $\delta$ can be extended to strings, that is, $\delta : Q \times \Sigma^* \rightarrow Q$, in the usual way. 
We use $\delta (q,s)!$ to denote that $\delta (q,s)$ is defined. The language {\em generated} by $G$ is the set of all strings defined in $G$ from the initial state $q_0$, that is,
$$
L(G) = \{ s \in \Sigma ^*: \delta (q_0,s)! \}.
$$
The language {\em marked} by $G$ is the set of all strings defined in $G$ from the initial state and end in a marked state:
$$
L_m(G) = \{ s \in L(G): \delta (q_o,s) \in Q_m \}.
$$

In general, a language $K \subseteq \Sigma ^*$ is a set of strings. For a string $s \in \Sigma ^*$, we use $s' \leq s$ to denote that $s'$ is a prefix of $s$. 
The (prefix) closure of $K$, denoted as $\overline{K}$, is the set of prefixes of strings in $K$, that is,
$$
\overline{K}=\{s':(\exists s\in K)s' \leq s\}
$$
A language is (prefix) closed if it equals its prefix closure. By the definition, $L(G)$ is closed. For a string $s \in \Sigma ^*$, we use $|s|$ to denote its length. For a set $x \subseteq Q$, we use $|x|$ to denote its cardinality (the number of its elements).

We use a controller, called supervisor, to control the system, called plant, so that some control objective is achieved. The supervisor can control some events and observe some other events. 
The set of events that can be controlled/disabled, called controllable events, is denoted by $\Sigma _c$ ($\subseteq \Sigma$). $\Sigma _{uc} = \Sigma - \Sigma _c$ is the set of uncontrollable events. 
The set of events that can be observed, called observable events, is denoted by $\Sigma _o$ ($\subseteq \Sigma$). $\Sigma _{uo} = \Sigma - \Sigma _o$ is the set of unobservable events. 
The set of observable transitions is denoted by $\delta_o = \{ (q, \sigma, q')\in \delta : \sigma \in \Sigma _o\}$; and the set of unobservable transitions is denoted by $\delta_{uo} = \{ (q, \sigma, q') \in \delta : \sigma \in \Sigma _{uo}\}$.

For a given string, its observation is described by the projection $P:\Sigma^* \rightarrow \Sigma^*_o$ which is defined as
\begin{equation*}
	\begin{aligned}
		& P(\varepsilon)= \varepsilon \\
		& P(\sigma) =\begin{cases}
			\sigma &\text{if $\sigma \in \Sigma_o$} \\
			\varepsilon &\text{if $\sigma\in \Sigma - \Sigma_o$} \\
		\end{cases}\\
		& P(s\sigma)=P(s)P(\sigma) ,  s\in \Sigma^*, \sigma \in \Sigma,
	\end{aligned}
\end{equation*}
where $\varepsilon$ is the empty string.

\subsection{ALTER Model}

Cyber attackers can change the readings of some vulnerable sensors. Hence, some observable events can be attacked. The terminology ``ALTER'' is used in this paper to name the model introduced in \cite{zheng2021modeling, zheng2023modeling}, which is reviewed below. For more detailed description of the ALTER model and its implementation, the reader is refereed to \cite{zheng2021modeling, zheng2023modeling, lin2023diagnosability, lin2024diagnosability, wang2024coordination, zheng2024opacity}.

Denote the set of observable events and transitions that may be attacked by $\Sigma _o^a \subseteq \Sigma _o$ and $\delta ^a = \{ (q, \sigma, q') \in \delta : \sigma \in \Sigma _o^a \}$, respectively. 
We call events $\sigma \in \Sigma ^a$ attackable events and transitions ${tr}=(q, \sigma, q')\in \delta ^a$ attackable transitions.
For an attackable transition $tr=(q, \sigma, q')\in \delta ^a$, we assume that an attacker can change the observation from  event $\sigma$ to any string in a language $A_{tr} \subseteq \Sigma_o ^*$. 
$A_{tr}$ is determined based the knowledge of the attacker and the corresponding sensor. We denote the set of all such attack languages as $\mathbb{A}$ 
$$
\mathbb{A}=\{A_{tr}:tr=(q, \sigma, q')\in \delta ^a\}.
$$

Sensor attacks are then modeled by a mapping from the set of transitions to be attacked to the set of languages $\mathbb{A}$ 
$$
\pi : \delta ^a\rightarrow \mathbb{A}
$$
where $\pi(tr)=A_{tr}$. Note that $\pi$ is transition-based. Note also that this general model allows for nondeterministic attacks and includes the following special cases. 
(1) No attack: if $\sigma \in A_{tr}$ and $\sigma$ is altered to $\sigma$ (no change), then there is no attack. 
(2) Deletion: if the empty string $\varepsilon \in A_{tr}$ and $\sigma$ is altered to $\varepsilon$, then $\sigma$ is deleted. 
(3) Replacement: if $\alpha \in A_{tr}$ and $\sigma$ is altered to $\alpha$, then $\sigma$ is replaced by $\alpha$. 
(4) Insertion: if $\sigma \alpha \in A_{tr}$ and $\sigma$ is altered to $\sigma \alpha$, then $\alpha$ is inserted.

If a string $s = \sigma _1 \sigma _2 ..., \sigma _{|s|} \in L(G)$ occurs in $G$, the set of possible strings after cyber attacks, denoted by $\Theta ^\pi(s)$, is obtained as follows. Let $q_k = \delta(q_0,\sigma _1\cdots\sigma _{k}), k= 1, 2, ..., |s|$, then
$$
\Theta ^\pi(s) = L_1 L_2 ... L_{|s|},
$$
where
\begin{align*}
	L_k = \left\{ \begin{array}{ll}
		\{ \sigma _k \} & \mbox{if } (q_{k-1}, \sigma _k, q_k) \not\in \delta ^a \\
		A_{(q_{k-1}, \sigma _k, q_k)} & \mbox{if } (q_{k-1}, \sigma _k, q_k) \in \delta ^a 
	\end{array} \right .
\end{align*}

Note that $\Theta ^\pi(s)$ may contain more than one string. Hence, $\Theta ^\pi$ is a mapping from $L(G)$ to $2^{\Sigma ^*}$:
$$
\Theta ^\pi: L(G) \rightarrow 2^{\Sigma ^*}.
$$

The observation under both partial observation and cyber attacks in the observation channel is then given by
$$
\Phi ^\pi = P \circ \Theta ^\pi,
$$
where $\circ$ denotes composition (of functions). In other words, for $s \in L(G)$, $\Phi ^\pi (s) = P ( \Theta ^\pi (s))$. Hence, $\Phi ^\pi$ is a mapping from $L(G)$ to $2^{\Sigma _o^*}$:
$$
\Phi ^\pi: L(G) \rightarrow 2^{\Sigma _o ^*}.
$$

We extend $P$, $\Theta ^\pi$, and $\Phi ^\pi$ from strings $s$ to languages $L \subseteq L(G)$ in the usual way as
\begin{equation} \label{Equation1}
	\begin{split}
		& P(L) = \{ w \in \Sigma^*_o: (\exists s \in L) w = P(s) \} \\
		& \Theta ^\pi(L) = \{ w \in \Sigma^*: (\exists s \in L) w  \in
		\Theta ^\pi(s) \} \\
		& \Phi ^\pi(L) = \{ w \in \Sigma^*_o: (\exists s \in L) w  \in \Phi ^\pi(s) \}.
	\end{split}
\end{equation}

The ALTER model is both general and specific. It is general in the sense that it includes deletion, insertion, and replacement as special cases. It is specific in the sense that the attacks are modeled by specific attack languages. 

\subsection{Supervisory Control}

After the occurrence of $s\in L(G)$, a supervisor $\mathcal{S}$ observes one of the string $w \in \Phi ^\pi (s)$. Based on $w$, $\mathcal{S}$ enables a set of events, denoted by $\mathcal{S}(w)$. Hence, $\cS$ is a mapping
$$
\mathcal{S} :\Phi ^\pi ( L(G)) \rightarrow \Gamma ,
$$
where $\Gamma = 2^{\Sigma}$ is the set of all possible controls. It is required that $\Sigma_{uc}\subseteq \mathcal{S}(w)$ because uncontrollable events cannot be disabled.

We assume that the disablements/enablements of some controllable events can be altered by attackers in the control channel from the supervisor to the plant in the sense that an attacker can enable an event that is disabled by the supervisor and/or disable an event that is enabled by the supervisor. 
Denote the set of controllable and may be attackable events by $\Sigma _c^a \subseteq \Sigma _c$. Note that uncontrollable events are always permitted to occur and the attackers cannot disable them.

Under cyber attacks in the control channel, for a given control $\gamma \in \Gamma$, some events in $\Sigma _c^a$ can be added to it or removed from it by attackers. 
Hence the possible controls after cyber attacks are
\begin{equation} \label{Equation11}
	\begin{split}
		\Delta(\gamma) = \{ \gamma_a \in \Gamma :  (\exists \gamma ', \gamma '' \subseteq  \Sigma _c^a ) \gamma_a = (\gamma - \gamma ')\cup  \gamma ''\} .
	\end{split}
\end{equation}

When the supervisor issues a control command $\mathcal{S}(w)$ after observing $w \in \Phi ^\pi ( L(G))$, it may be altered under cyber attacks. We use $\cS^a (w)$ to denote the set of all possible control commands that may be received by the plant under cyber attacks, that is,
\begin{equation} \label{Equation12}
	\cS^a(w)=\Delta(\cS(w)).
\end{equation}

We denote the supervised system under attacks as $\mathcal{S} ^a/G$. Because of the attacks, the language generated by the supervised system, denoted by $L(\mathcal{S} ^a/G)$, is nondeterministic, and hence may not be unique. 
We consider the upper bound of all languages that can be generated by the supervised system $\mathcal{S} ^a/G$. We call the upper bound {\em large language}, which is defined below.

\begin{definition} \label{Definition1} 
	The large language generated by the supervised system	$L(\mathcal{S} ^a/G)$, denoted by $L_a (\mathcal{S} ^a/G)$, is 	defined recursively as follows.
	
	\begin{enumerate}
		\item
		The empty string belongs to $L_a (\cS ^a/G)$, that is,
		$$
		\varepsilon \in L_a (\cS ^a/G).
		$$
		\item
		If $s$ belongs to $L_a (\cS ^a/G)$, then for all $\sigma \in \Sigma$, $s \sigma$ belongs to $L_a (\cS ^a/G)$ if and only if $s \sigma$ is physically possible in $L(G)$ and $\sigma$ is uncontrollable or enabled by $\cS ^a$ in {\em some} situations, that is, for all $ s \in L_a (\cS ^a/G)$ and $\sigma \in \Sigma$,
		\begin{equation} \label{Equation9}
			\begin{split}
				&  s \sigma\in L_a (\cS ^a/G) \\
				\Leftrightarrow & s \sigma \in L(G) \wedge (\sigma \in \Sigma _{uc} \\
				& \vee (\exists w \in \Phi ^\pi (s))(\exists \gamma_a \in \cS ^a(w)) \sigma \in \gamma_a).
			\end{split}
		\end{equation}
		
	\end{enumerate}
	
\end{definition}

Our goal is to control the system $G$ to be safe under cyber attacks. The safety specification is described by a (closed) specification language $K$ generated by a sub-automaton of $G$ as
\begin{align*}
	H = (Q_H, \Sigma, \delta_H, q_0),
\end{align*}
where $Q_H \subseteq Q$ and $\delta_H = \delta|_{Q_H \times \Sigma} \subseteq \delta$. Intuitively, $Q_H$ is the set of safe states and $Q-Q_H$ is the set of unsafe states.

To achieve this goal, a supervisor must be designed such that the large language of the supervised system $L_a(\mathcal{S}^a/G)$ is equal to or contained in the specification language $K$. 

To this end, let us consider the set of possible states that $\mathcal{S}^a/G$ may be in, after observing a string $w \in \Phi ^\pi (L(H))$. This set is called the state estimate and is defined as
\begin{align*}
	SE^\pi_H(w)= \{& q \in Q: (\exists s \in L(H)) \\
	& w \in \Phi ^\pi (s) \wedge \delta  (q_0, s)=q \}.
\end{align*}
Note that the states that may be reached by $\mathcal{S}^a/G$ are those in $H$, since we want $L_a(\cS^a/G) \subseteq K =L(H)$.

To compute state estimate $SE^\pi_H(w)$, we proceed as follows. For each attackable transition $tr\in \delta^a$, let us assume that $A_{tr}$ is marked by an automaton 
$$
F_{tr} = (Q_{tr} , \Sigma, \delta_{tr}, q_{0,tr}, Q_{m,tr}).
$$
Note that not all states in $F_{tr}$ are marked. We assume that $F_{tr}$ is trim, that is, all states are accessible from $q_{o,{tr}}$ and co-accessible to $Q_{m,{tr}}$.

We replace a transition $tr=(q, \sigma, q') \in \delta _H^a = \delta _H \cap \delta ^a$ in $H$ by $F_{tr}$ as follows.
\begin{align*}
	H_{tr \rightarrow F_{tr}} = ( Q_H \cup Q_{tr}, \Sigma , \delta_{tr \rightarrow F_{tr}}, q_0 )
\end{align*}
where $\delta_{tr \rightarrow F_{tr}} = (\delta_H - \{ (q, \sigma, q') \} ) \cup \delta_{tr} \cup \{ (q, \varepsilon, q_{0,tr}) \} \cup \{ (q_{m,tr}, \varepsilon, q'): q_{m,tr} \in Q_{m,tr} \}$. 
Note that if $F_{tr}$ has only one marked state, that is, $Q_{m,tr}=\{ q_{m,tr} \}$, then we can use a shortcut by merging $q$ and $q'$ with $q_{0,tr}$ and $q_{m,tr}$, respectively, without the $\varepsilon$-transitions $(q, \varepsilon, q_{0,tr})$ and $(q_{m,tr}, \varepsilon, q')$.


Denote the automaton after replacing all transitions subject to attacks in $H$ as
$$
H^\diamond=( Q^\diamond, \Sigma , \delta_H^ \diamond , q_0, Q^\diamond_m)=( Q_H \cup \hat{Q}, \Sigma , \delta_H^ \diamond , q_0, Q_H),
$$
where $\hat{Q}= \cup_{tr \in \delta _H^a} Q_{tr}$ is the set of states added during the replacement and $Q^\diamond_m=Q_H$ is the set of marked states. Note that $H^\diamond$ is a nondeterministic automaton, 
that is, $\delta_H^ \diamond$ is a mapping $\delta_H^ \diamond : Q^\diamond \times \Sigma \rightarrow 2^{Q^\diamond}$. Clearly, $H^\diamond$ marks $\Theta ^\pi (L(H))$:
$$
\Theta ^\pi (L(H)) = L_m(H^\diamond),
$$	

To model the partial observation, we replace unobservable transitions by $\varepsilon$-transitions and denote the resulting automaton as
$$
H_{\varepsilon}^\diamond=( Q_H \cup \hat{Q}, \Sigma_o , \delta _{H, \varepsilon}^\diamond , q_0, Q_H)
$$
where $\delta _{H, \varepsilon}^\diamond = \{ (q, \sigma, q'): (q, \sigma, q') \in \delta ^\diamond \wedge \sigma \in \Sigma _o \} \cup \{ (q, \varepsilon, q'): (q, \sigma, q') \in \delta ^\diamond \wedge \sigma \not\in \Sigma _o \} $. 
$H^\diamond_\varepsilon$ is also a nondeterministic automaton.

The set of strings that can be observed under partial observation and cyber attacks in observation channel, $\Phi ^\pi (L(H))$, is given by
$$
\Phi ^\pi (L(H)) = L_m(H_{\varepsilon}^\diamond),
$$	
where $L_m(H_{\varepsilon}^\diamond)= \{ s \in \Sigma_o^* : \delta _{H, \varepsilon}^\diamond (q_0, s) \cap Q_H \not= \emptyset \}$ is the language marked by $H_{\varepsilon}^\diamond$.

To obtain the state estimates, we convert $H_{\varepsilon} ^\diamond$ to a deterministic automaton $H_{obs}^\diamond$, called CA-observer, as follows.
\begin{equation}
	\begin{aligned}
		H_{obs}^\diamond & =(X,\Sigma _o, \xi, x_0, X_m) \\
		&= Ac(2^{Q_H \cup \hat{Q}},\Sigma _o, \xi, UR(\{ q_0\}), X_m), \label{OBS}  \notag
	\end{aligned}
\end{equation}
where $Ac(.)$ denotes the accessible part; $UR(.)$ is the unobservable reach defined, for $x \subseteq Q_H \cup \hat{Q}$, as
$$
UR(x) = \{ q \in Q_H \cup \hat{Q}: (\exists q' \in x) q\in \delta _{\varepsilon}^\diamond (q', \varepsilon) \}.
$$

The transition function $\xi$ is defined, for $x \in X$ and $\sigma \in \Sigma _o$ as
$$
\xi (x, \sigma) = UR(\{q \in Q_H \cup \hat{Q}: (\exists q' \in x)q\in \delta _{H, \varepsilon}^\diamond (q',\sigma) \}).
$$
The marked states are defined as
$$
X_m = \{x \in X: x \cap Q_H \not = \emptyset \}.
$$
It is well-known that $L(H_{\varepsilon}^\diamond) =L(H_{obs} ^\diamond)$ and $L_m(H_{\varepsilon}^\diamond) =L_m(H_{obs} ^\diamond)$. Hence, $\Phi ^\pi (L(H)) = L_m(H_{obs}^\diamond)$.
The following proposition is proved in \cite{zheng2023modeling},

\begin{proposition}	\label{PropositionSE} 	
	After observing $w \in \Phi ^\pi (L(H))$, the state estimate $SE^\pi_H(w)$ is given by
	\begin{equation} \label{SE}
		\begin{split}
			SE^\pi_H(w)= \xi (x_0, w) \cap Q_H.
		\end{split}
	\end{equation}
\end{proposition} 

Using state estimates, we can construct a state-estimate-based supervisor $\cS _{CA} : \Phi ^\pi ( L(G)) \rightarrow \Gamma$, called CA-supervisor, that ensure the supervised system will not enter unsafe states in $Q-Q_H$ as follows. 
For $w \in \Phi ^\pi ( L(G))$,
\begin{equation} \label{SCA}
	\begin{split}
		\cS _{CA} (w) =&  \{\sigma \in \Sigma : (\forall q \in SE^\pi_H(w))q \in Q_H \\
		&\wedge \delta (q,\sigma) \in Q \Rightarrow \delta (q,\sigma) \in Q_H \}
	\end{split}
\end{equation}

To find the existence condition for a supervisor $\cS$ such that $L_a(\mathcal{S}^a/G)=K$, we extend controllability and observability to CA-controllability and CA-observability as follows \cite{zheng2021modeling, zheng2023modeling}.

\begin{definition} \label{Definition2} 
	A closed language $K \subseteq L(G)$ is {\em CA-controllable} with respect to $L(G)$, $\Sigma _{uc}$, and $\Sigma ^a_c$ if
	$$ \label{C}
	K (\Sigma _{uc} \cup \Sigma ^a_c) \cap L(G) \subseteq K. \notag
	$$
\end{definition}

\begin{definition} \label{Definition3} 
	A closed language $K \subseteq L(G)$ is {\em CA-observable} with respect to $L(G)$, $\Sigma _{o}$, $\Sigma ^a_o$ and $\Phi^\pi$ if
	\begin{align*}
		& (\forall s \in \Sigma ^*)(\forall \sigma \in \Sigma) (s \sigma \in K \\
		& \Rightarrow (\exists w \in \Phi ^\pi (s))(\forall s' \in (\Phi ^\pi)^{-1} (w)) \\
		& (s' \in K \wedge s' \sigma \in L(G) \Rightarrow s' \sigma \in K)),
	\end{align*}
	where $(\Phi ^\pi)^{-1}$ is the inverse mapping of $\Phi ^\pi$: for $w \in \Phi ^\pi (L(G))$,
	\begin{equation} \label{Phi}
		\begin{split}
			(\Phi ^\pi)^{-1} (w) = \{ s' \in L(G) : w \in \Phi ^\pi (s') \}.
		\end{split}
	\end{equation}
\end{definition}

With CA-controllability and CA-observability defined, we have the
following theorem, which is proved in \cite{zheng2023modeling}.

\begin{theorem} \label{TheoremEC} 	
	Consider a discrete event system $G$ under cyber attacks. For a nonempty closed language $K \subseteq L(G)$ generated by a sub-automaton $H \sqsubseteq G$, 
	there exists a supervisor $\cS :\Phi ^\pi ( L(G)) \rightarrow \Gamma$ such that $L_a(\cS^a/G)=K$ if and only if $K$ is CA-controllable with respect to $L(G)$, $\Sigma _{uc}$, and $\Sigma ^a_c$ and CA-observable with respect to $L(G)$, $\Sigma _{o}$, $\Sigma ^a_o$ and $\Phi^\pi$. 
	Furthermore, if CA-controllability and CA-observability are satisfied, then CA-supervisor $\cS_{CA}$ defined in Equation (\ref{SCA}) is such a supervisor, that is,		
	$$
	L_a(\cS^a_{CA}/G)=K.
	$$
\end{theorem} 


\section{Checking CA-observability} \label{s3}

It is clear that $K$ is {CA-controllable} with respect to $L(G)$, $\Sigma _{uc}$, and $\Sigma ^a_c$ if and only if $K$ is {controllable} with respect to $L(G)$ and $\Sigma _{uc} \cup \Sigma ^a_c$. 
In this sense, CA-controllability can be ``converted'' to (conventional) controllability. Therefore, there is no need to investigate CA-controllability further, as all the results on controllability can be used. 
For example, checking CA-controllability can be done using the exists algorithms for controllability. However, CA-observability cannot be ``converted'' to (conventional) observability. Therefore, we need to investigate CA-observability. In this section, we develop a method to check CA-observability. 

Given a language $K=L(H)$, we want to check if $K$ is CA-observable (with respect to $L(G)$, $\Sigma _{o}$, $\Sigma ^a_o$ and $\Phi^\pi$) or not.
To focus on CA-observability, let us assume, without loss of generality, that $K$ is CA-controllable (with respect to $L(G)$, $\Sigma _{uc}$, and $\Sigma ^a_c$), which can be ensured if all events are controllable and there are no cyber attacks in the control channel, that is, $\Sigma _c = \Sigma$ and $\Sigma _c^a = \emptyset$.

%
%
%

For $w \in \Phi ^\pi (L(G))$, define the set of events to be enabled in state estimate $SE^\pi_H(w)$ to achieve the large language as
\begin{align*}
	\mu(SE^\pi_H(w)) =&\{\sigma \in \Sigma: (\forall q \in SE^\pi_H(w))q \in Q_H \\
	&\wedge \delta (q,\sigma) \in Q \Rightarrow \delta (q,\sigma) \in Q_H \}.
\end{align*}
Note that $\mu$ is a function $\mu: 2^Q \rightarrow \Sigma$.

From the definitions of $\cS _{CA} (w)$ and $\mu(SE^\pi_H(w))$, it is clear that
\begin{equation} \label{equL}
	\begin{split}
		\cS _{CA} (w) = & \{\sigma \in \Sigma : (\forall q \in SE^\pi_H(w))q \in Q_H \\
		& \wedge \delta (q,\sigma) \in Q \Rightarrow \delta (q,\sigma) \in Q_H \} \\
		= & \mu(SE^\pi_H(w)) .
	\end{split}
\end{equation}

Since we assume that $K$ is CA-controllable, we have, for $w \in \Phi ^\pi (L(G))$,
\begin{equation} \label{Eq9}
	\begin{split}
		\Sigma _{uc} \cup \Sigma ^a_c \subseteq \cS _{CA} (w) = \mu(SE^\pi_H(w)).
	\end{split}
\end{equation}

The augmented automaton is then defined as 
$$
H_{aug} = (Z, \Sigma,\rho, z_0) = Ac(Q \times {2^{X}}, \Sigma, \rho, z_0).
$$
A state of $H_{aug}$ is denoted by $z=(q, \mathcal{X})$, where $q \in Q$ is the current state of $G$ and $\mathcal{X} \subseteq 2^X$ is the set of possible state estimates. The initial state is $z_0 = (q_0, \{x_0\})$.

We extend the definition of $\Phi ^\pi$ to strings starting at any state $q \in Q$ by treating $q$ as the initial state,
$$
\Phi ^\pi: Q \times \Sigma^* \rightarrow 2^{\Sigma_o ^*}.
$$
If $q = q_0$, we still denote $\Phi ^\pi(q_0, s)$ by $\Phi ^\pi(s)$.

For $z = (q, \mathcal{X})$ and $\sigma \in \Sigma$,
\begin{align} \label{rho1}
	\rho(z,\sigma)! \Leftrightarrow \delta(q,\sigma)! \wedge(\exists x \in \mathcal{X}) \sigma \in \mu(x \cap Q_H).
\end{align}

If $\rho(z,\sigma)$ is defined, then $\rho(z,\sigma) = z' = (q', \mathcal{X'})$ is given as follows.
\begin{equation} \label{rho2}
	\begin{split}
		q'= & \delta(q, \sigma) \\
		\mathcal{X'} = & \{x' \in X: (\exists x \in \mathcal{X}) \\
		& (\exists v \in \Phi ^\pi(q, \sigma)) x' = \xi(x, v)\} .
	\end{split}
\end{equation}

The augmented automaton can be used to determine the large language as shown in the following theorem.

\begin{theorem}	\label{theoaug}
	The language generated by automaton $H_{aug}$ is equal to $L_a({\cS^a_{CA}/G})$, that is
	$$
	L(H_{aug}) = L_a({\cS^a_{CA}/G}).
	$$	
\end{theorem}

By Theorem \ref{theoaug}, we can check CA-observability as follows.

\begin{theorem}	\label{theoaug4}
	Let $K=L(H)$. $K$ is CA-observable with respect to $L(G)$, $\Sigma _{o}$, $\Sigma ^a_o$ and $\Phi^\pi$ if and only if $L(H_{aug}) = K$ or equivalently
	$$
	(\forall z=(q, \mathcal{X}) \in Z) \Gamma_{H_{aug}}(z) = \Gamma_H(q),
	$$
	where $\Gamma_{H_{aug}}(z) = \{\sigma \in \Sigma: \rho(z, \sigma) \rm ~is~ defined\}$ and $\Gamma_H(q)$ $=$ $\{\sigma \in \Sigma: \delta_H (q, \sigma) \rm ~is~ defined\}$.
	
\end{theorem}

\begin{algorithm}[hbt!] \label{alg1}
	
	\caption{(Check CA-observability of $K$)}
	\begin{algorithmic}[1]
		
		\REQUIRE $G, H, \Sigma_o, \Sigma _o^a, \mathbb{A}, \pi$;
		
		\ENSURE $\rm CAO$;
		
		\STATE Calculate the nondeterministic automaton 
		\begin{align*}
			H^\diamond & =( Q^\diamond, \Sigma , \delta_H^ \diamond , q_0, Q^\diamond_m) \\
			& =( Q_H \cup \hat{Q}, \Sigma , \delta_H^ \diamond , q_0, Q_H);
		\end{align*}
		
		\STATE Calculate the CA-observer 
		\begin{align*}
			H_{obs}^\diamond & =(X,\Sigma _o, \xi, x_0, X_m) \\
			&= Ac(2^{Q_H \cup \hat{Q}},\Sigma _o, \xi, UR(\{ q_0\}), X_m);
		\end{align*}

		\STATE Calculate the augmented automaton 
		\begin{align*}
			H_{aug} & = (Z, \Sigma,\rho, z_0) \\
			& = Ac(Q \times {2^{X}}, \Sigma, \rho, z_0);
		\end{align*}
		
		\STATE If $(\forall z=(q, \mathcal{X}) \in Z) \Gamma_{H_{aug}}(z) = \Gamma_H(q)$, then 
		
		$\rm CAO = Yes$;
		
		else
		
		$\rm CAO = No$;
		
		\STATE End.
		
	\end{algorithmic}

\end{algorithm}

The steps to check CA-observability (with respect to $L(G)$, $\Sigma _{o}$, $\Sigma ^a_o$ and $\Phi^\pi$) is summarized in Algorithm 1. $K$ is CA-observable if and only if the output of the algorithm $\rm CAO = Yes$.

The computational complexity of Algorithm 1 can be obtained as follows. The computational complexity of Step 1 is $O(| Q_H \cup \hat{Q} |\cdot |\Sigma|)$. 
The computational complexity of Step 2 is $O( 2^{| Q_H \cup \hat{Q} |} \cdot |\Sigma _o| )$. 
The computational complexity of Step 3 is $O( 2^{2^{| Q_H \cup \hat{Q} |}} \cdot| Q_H \cup \hat{Q} | \cdot |\Sigma| )$. 
The computational complexity of Step 4 is also $O( 2^{2^{| Q_H \cup \hat{Q} |}} \cdot| Q_H \cup \hat{Q} | \cdot |\Sigma| )$. 
Hence, the computational complexity of Algorithm 1 is $O( 2^{2^{|Q_H \cup \hat{Q}|}} \cdot| Q_H \cup \hat{Q} | \cdot |\Sigma| )$.

Let us illustrate Algorithm 1 in the following example.

\begin{example}	
	Let us consider the discrete event system $G$ in Fig. \ref{fig:G}. In the figures, symbol $\rightarrow$ denotes the initial state and double circles denote marked states. Assume that all events are observable, that is, $\Sigma = \Sigma_o = \{\alpha, \beta, \gamma, \lambda \}$. 
	
	\begin{figure}[hbt!]
		\centering
		\includegraphics[scale=0.9]{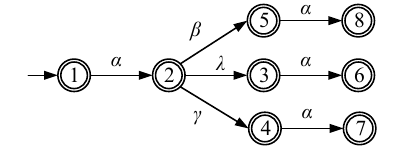}
		\caption{Discrete event system $G$ of Example 1.  }
		\label{fig:G}
	\end{figure}

	Assume that $\lambda$ may be attacked, that is, $\Sigma_o^a = \{\lambda \}$. For transition $tr=(2, \lambda, 3) \in \delta^a$, the attack language $F_{tr}$ is marked by the automaton shown in Fig. \ref{fig:Ftr}.

	\begin{figure}[hbt!]
		\centering
		\includegraphics[scale=0.9]{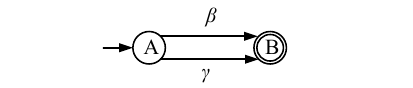}
		\caption{Attack language $F_{tr}$ for $tr=(2, \lambda, 3)$ of Example 1.  }
		\label{fig:Ftr}
	\end{figure}

	Let the specification language $K$ be generated by the automaton shown in Fig. \ref{fig:K1}.

	\begin{figure}[hbt!]
		\centering
		\includegraphics[scale=0.9]{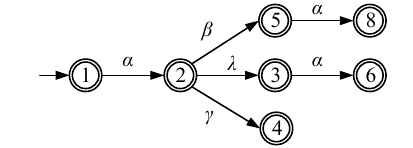}
		\caption{Specification language $K$ of Example 1.  }
		\label{fig:K1}
	\end{figure}

	By Step 1 of Algorithm 1, the nondeterministic automaton $H^\diamond$ after replacing all transitions subject to attacks in $H$ is calculated as shown in Fig. \ref{fig:H1}.
	
	\begin{figure}[hbt!]
		\centering
		\includegraphics[scale=0.9]{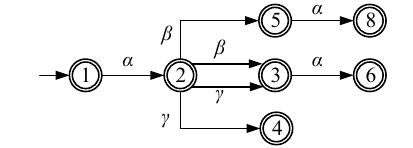}
		\caption{Nondeterministic automaton $H^\diamond$ of Example 1.  }
		\label{fig:H1}
	\end{figure}

	By Step 2 of Algorithm 1, the CA-observer $H_{obs}^\diamond$ is calculated as shown in Fig. \ref{fig:Ho1}.
	
	\begin{figure}[hbt!]
		\centering
		\includegraphics[scale=0.9]{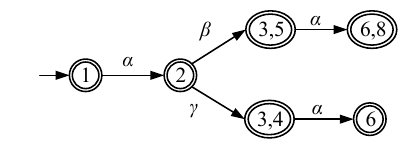}
		\caption{CA-observer $H_{obs}^\diamond$ of Example 1.  }
		\label{fig:Ho1}
	\end{figure}

	By Step 3 of Algorithm 1, the augmented automaton $H_{aug}$ is calculated as shown in Fig. \ref{fig:Ga1}. 
	
	\begin{figure}[hbt!]
		\centering
		\includegraphics[scale=0.9]{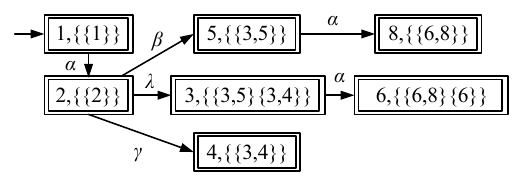}
		\caption{Augmented automaton $H_{aug}$ of Example 1.  }
		\label{fig:Ga1}
	\end{figure}

	By Step 4 of Algorithm 1, $(\forall z=(q, \mathcal{X}) \in Z) \Gamma_{H_{aug}}(z) = \Gamma_H(q)$ is true. Therefore, $K$ is CA-observable.

\end{example}

\section{Properties of CA-observability} \label{s4}

Since CA-controllability can be ``converted'' to (conventional) controllability, CA-controllability naturally inherits many ``good" proerties of controllability such as CA-controllability is closed under union and intersection. However, this inheritance does not hold for CA-observability. Hence, in this section, we investigate properties of CA-observability, that is, whether CA-observability is closed under (set) intersection or union. Recall that observability is closed under intersection, but not under union \cite{lin1988observability}. 

\begin{theorem} \label{Proposition3}
	CA-observability is not closed under (set) intersection, that is, the fact that $K_1$ and $K_2$ are CA-observable (with respect to $L(G)$, $\Sigma _{o}$, $\Sigma ^a_o$ and $\Phi^\pi$) 
	does not imply that $K_1 \cap K_2$ is CA-observable (with respect to $L(G)$, $\Sigma _{o}$, $\Sigma ^a_o$ and $\Phi^\pi$).
	
\end{theorem}

Since observability is closed under intersection, there are clear difference between observability and CA-observability. 
In particular, while the infimal observable superlanguage of $K$ always exist, the infimal CA-observable superlanguage of $K$ may not exist.

\begin{theorem} \label{Proposition2}	
	CA-observability is not closed under (set) union, that is, the fact that $K_1$ and $K_2$ are CA-observable (with respect to $L(G)$, $\Sigma _{o}$, $\Sigma ^a_o$ and $\Phi^\pi$) 
	does not imply that $K_1 \cup K_2$ is CA-observable (with respect to $L(G)$, $\Sigma _{o}$, $\Sigma ^a_o$ and $\Phi^\pi$).	
\end{theorem}

\section{CA-controllable and CA-observable sublanguages} \label{s5}

Since CA-observability is not closed under union, the supremal CA-observable sublanguage of $K$ may not exist. Hence, the supremal CA-controllable and CA-observable sublanguage of $K$ also may not exsit. 
In the following, we propose to calculate a CA-controllable and CA-observable sublanguage of $K$ that can be used to synthesize a supervisor to ensure safety/legality of the supervised system under cyber attacks when given $K$ is not CA-controllable and/or CA-observable. We expect the sublanguage to be as large as possible.

Since CA-controllability is closed under union, the supremal CA-controllable sublanguage of $K$ with respect to $L(G)$, $\Sigma _{uc}$, and $\Sigma ^a_c$, denoted by $K^\uparrow$, exists. By the well-known result in supervisory control \cite{cassandras2021introduction}, $K^\uparrow$ is generated by 
\begin{align*}
	H^\uparrow = & (Q_{H}^\uparrow, \Sigma, \delta_{H}^\uparrow, q_{o}),
\end{align*}
where $Q_{H}^\uparrow$ is obtained by removing states in $H$ that can reach illegal states $Q-Q_{H}$ via a string of uncontrollable and/or attackable events in $G$. In other words,
\begin{align*}
	Q_{H}^\uparrow = & Q_{H} - \{q \in Q_{H} : (\exists s \in (\Sigma _{uc} \cup \Sigma ^a_c)^*) \\
	& \delta (q, s) \in Q-Q_{H} \} \\
	\delta_{H}^\uparrow = & \delta|_{Q_{H}^\uparrow \times \Sigma} .
\end{align*}

After obtaining $H^\uparrow$, we construct automaton $H_{aug} ^\uparrow$ in the same way as the construction of automaton $H_{aug}$. The entire procedure is summarized in Algorithm 2.

\begin{algorithm}[hbt!] \label{alg2}
	
	\caption{(Construct automaton $H_{aug} ^\uparrow$)}
	\begin{algorithmic}[1]
		
		\REQUIRE $G, H, \Sigma_o, \Sigma _o^a, \mathbb{A}, \pi$;
		
		\ENSURE $H_{aug} ^\uparrow$;
		
		\STATE Calculate the automaton 
		$$
		{H^\uparrow} = (Q_{H}^\uparrow, \Sigma, \delta_{H}^\uparrow, q_{o});
		$$
		
		\STATE Calculate the automaton 
		\begin{align*}
			H^{\uparrow \diamond} & =( Q^{\uparrow \diamond}, \Sigma , \delta_H^{\uparrow \diamond} , q_0, Q^{\uparrow \diamond} _m) \\
			& =( Q_H^\uparrow \cup \hat{Q}^\uparrow, \Sigma , \delta_H^{\uparrow \diamond} , q_0, Q_H^\uparrow);
		\end{align*}
	
		\STATE Calculate the automaton 
		\begin{align*}
			H_{\varepsilon}^{\uparrow \diamond} & =( Q_H^\uparrow \cup \hat{Q}^\uparrow, \Sigma_o , \delta _{H, \varepsilon}^{\uparrow \diamond} , q_0, Q_H^\uparrow);
		\end{align*}
		
		\STATE Calculate the CA-observer 
		\begin{align*}
			H_{obs} ^{\uparrow \diamond} & =(X ^\uparrow,\Sigma _o, \xi ^\uparrow, x_0, X_m ^\uparrow) \\
			&= Ac(2^{Q_H^\uparrow \cup \hat{Q} ^\uparrow},\Sigma _o, \xi ^\uparrow, UR(\{ q_0\}), X_m ^\uparrow);
		\end{align*}
		
		\STATE Calculate the augmented automaton 
		\begin{align*}
			H_{aug}^\uparrow & = (Z^\uparrow, \Sigma, \rho^\uparrow, z_0^\uparrow) \\
			& = Ac(Q \times {2^{X}}^\uparrow, \Sigma, \rho^\uparrow, z_0^\uparrow);
		\end{align*}
		
		\STATE End.
		
	\end{algorithmic}
	
\end{algorithm}

We have the following result regarding the language $L(H_{aug} ^\uparrow)$.

\begin{theorem}	\label{Theorem6}	
	The language generated by automaton $H_{aug} ^\uparrow$, $L(H_{aug} ^\uparrow)$, is a CA-controllable (with respect to $L(G)$, $\Sigma _{uc}$, and $\Sigma ^a_c$) and CA-observable (with respect to $L(G)$, $\Sigma _{o}$, $\Sigma ^a_o$ and $\Phi^\pi$) sublanguage of $K$.
	
\end{theorem}

The computational complexity of Algorithm 2 can be obtained as follows. The computational complexity of Step 1 is bounded by $O(| Q_H |\cdot |\Sigma|)$. 
The computational complexity of Step 2 is bounded by $O(| Q_H \cup \hat{Q} |\cdot |\Sigma|)$. 
The computational complexity of Step 3 is also bounded by $O(| Q_H \cup \hat{Q} |\cdot |\Sigma|)$. 
The computational complexity of Step 4 is bounded by $O( 2^{| Q_H \cup \hat{Q} |} \cdot |\Sigma _o| )$. 
The computational complexity of Step 5 is bounded by $O( 2^{2^{| Q_H \cup \hat{Q} |}} \cdot| Q_H \cup \hat{Q} | \cdot |\Sigma| )$. 
Hence, the computational complexity of Algorithm 2 is bounded by $O( 2^{2^{| Q_H \cup \hat{Q} |}} \cdot| Q_H \cup \hat{Q} | \cdot |\Sigma| )$.

Note that if $K$ is CA-controllable and CA-observable, we then have $K=K^{\uparrow}$ and then $\cS_{CA}^{\uparrow}$ is equal to $\cS_{CA}$.

In order to enable as many events as possible, for any observation $w$, supervisor $\cS_{CA}^{\uparrow}$ only disable those events which drive states in $SE^\pi_{H^\uparrow} (w)$ into illegel states. However, since $SE^\pi_{H^\uparrow} (w)$ is not equal to  $SE^\pi_{\cS_{CA}^{\uparrow a}/G}(w)$, $\cS_{CA}^\uparrow$ may disable more events than necessary and hence $L_a({\cS_{CA}^{\uparrow a}/G})$ may be not maximal. How to find a maximal CA-controllable and CA-observable sublanguage is still an open problem.



\section{Conclusion} \label{s6}

In this paper, we investigate CA-observability. The main contributions are summarized as follows. 1. By constructing an augmented automaton which integrates the information of the given discrete event system and the state estimates under cyber attacks, we propose an appraoch to check CA-observability. 2. We investigate the properties of CA-observability which is not closed  under (set) intersection or under union. 3. By calculating the supremal CA-controllable sublanguage and then control it with a state-estimate-based supervisor, we obtain a CA-controllable and CA-observable sublanguage and propose a state-esitamte-based supevisor to constraint the dynamics of the closed-loop system in the sublanguage. 

In the future, we plan to extend these results to distributed discrete event systems and the nonblocking supervisory control problem under cyber attacks. We will also investigate algorithms to construct a maximally-permissive supervisor. 

\vspace{-5 mm}
\appendices 
\section{PROOF OF THEOREM \ref{theoaug}}
	For $s \in L(H_{aug})$, denote $\rho(z_0, s) = z = (q, \mathcal{X})$. Let us first prove
	\begin{equation} \label{rho3}
		\begin{split}
			q &= \delta(q_0, s) \\
			\mathcal{X} &= \{x \in X: (\exists w \in \Phi ^\pi (s)) x = \xi(x_0, w) \} 
		\end{split}
	\end{equation}
	by induction on the length $|s|$ of $s$.
	
	{\em Induction Base:}  Since the initial state is $z_0 = (q_0, \{x_0\})$, for $|s| = 0$, that is, $s=\varepsilon$, we have
	\begin{align*}
		q = & q_0 = \delta(q_0, \varepsilon) \\
		\mathcal{X} = & \{x_0\} = \{ \xi(x_0, \varepsilon) \} \\
		= & \{x \in X: (\exists w \in \Phi ^\pi (\varepsilon)) x = \xi(x_0, w) \} \\
		& (\mbox{because } \Phi ^\pi (\varepsilon) = \{ \varepsilon \}) .
	\end{align*}
	
	{\em Induction Hypothesis:} Assume that for all $s \in \Sigma^*$, $|s| \leq m$, $\rho(z_0, s) = z = (q, \mathcal{X})$
	\begin{align*}
		q &= \delta(q_0, s) \\
		\mathcal{X} &= \{x \in X: (\exists w \in \Phi ^\pi (s)) x = \xi(x_0, w) \} .
	\end{align*}
	
	{\em Induction Step:} We show that for all $s \in \Sigma^*$, $\sigma \in \Sigma$, $|s\sigma| = m+1$, $\rho(z_0, s \sigma) = z' = (q', \mathcal{X'})$
	\begin{align*}
		q' &= \delta(q_0, s \sigma) \\
		\mathcal{X'} &= \{x \in X: (\exists w' \in \Phi ^\pi (s \sigma)) x = \xi(x_0, w') \} .
	\end{align*}
	as follows. 
	\begin{align*}
		q' = & \delta(q, \sigma) \\
		& (\mbox{by Equation (\ref{rho2})}) \\
		= & \delta(\delta(q_0, s), \sigma) \\
		& (\mbox{by Induction Hypothesis}) \\
		= & \delta(q_0, s\sigma) .
	\end{align*}
	\begin{align*}
		\mathcal{X'} = & \{x' \in X: (\exists x \in \mathcal{X}) \\
		& (\exists v \in \Phi ^\pi(q, \sigma)) x' = \xi(x, v)\} \\
		& (\mbox{by Equation (\ref{rho2})}) \\
		= & \{x' \in X: (\exists w \in \Phi ^\pi (s)) x = \xi(x_0, w) \\
		& \wedge (\exists v \in \Phi ^\pi(q, \sigma)) x' = \xi(x, v)\} \\
		& (\mbox{by Induction Hypothesis}) \\
		= & \{x' \in X: (\exists w \in \Phi ^\pi (s)) x = \xi(x_0, w) \\
		& \wedge (\exists v \in \Phi ^\pi(\delta(q_0, s), \sigma)) x' = \xi(\xi(x_0, w), v)\} \\
		& (\mbox{because } q = \delta(q_0, s)) \\
		= & \{x' \in X: (\exists w \in \Phi ^\pi (s)) \\
		& (\exists v \in \Phi ^\pi(\delta(q_0, s), \sigma)) x' = \xi(\xi(x_0, w), v)\} \\
		= & \{x' \in X: (\exists w \in \Phi ^\pi (s)) \\
		& (\exists v \in \Phi ^\pi(\delta(q_0, s), \sigma)) x' = \xi(x_0, w v)\} \\
		= & \{x \in X: (\exists w v \in \Phi ^\pi (s\sigma)) x = \xi(x_0, w v) \} \\
		& (\mbox{by the definition of } \Phi ^\pi ) \\
		= & \{x \in X: (\exists w' \in \Phi ^\pi (s\sigma)) x = \xi(x_0, w') \} 
		\vspace{-2 mm}
	\end{align*}
\vspace{-2 mm}
	Let us now prove that, for  $s \in L(H_{aug})$, $\sigma \in \Sigma$, $s \sigma \in L(G)$,
	\begin{equation} \label{rho4}
		\begin{split}
			& \rho(\rho(z_0, s),  \sigma)! \\
			\Leftrightarrow 
			& s \sigma \in L(G)  \wedge  (\exists w \in \Phi ^\pi (s)) \sigma \in \mu(\xi(x_0, w) \cap Q_H ).
		\end{split}
	\end{equation}
		\vspace{-5 mm}
		
	Indeed,
		\vspace{-2 mm}
	\begin{align*}
		&\rho(\rho(z_0, s),  \sigma)!\\
		\Leftrightarrow 
		& \rho(z_0, s) = z = (q, \mathcal{X}) \wedge \rho(z,\sigma)! \\
		\Leftrightarrow 
		& \rho(z_0, s) = z = (q, \mathcal{X}) \wedge \delta(q,\sigma)! \wedge (\exists x \in \mathcal{X}) \sigma \in \mu(x \cap Q_H) \\
		& (\mbox{by Equation (\ref{rho1})}) \\
		\Leftrightarrow 
		& \rho(z_0, s) = z = (q, \mathcal{X}) \wedge \delta(q,\sigma)! \wedge (\exists w \in \Phi ^\pi (s)) \\
		& x = \xi(x_0, w) \wedge \sigma \in \mu(x \cap Q_H) \\
		& (\mbox{by Equation (\ref{rho3})}) \\
		\Leftrightarrow 
		& \rho(z_0, s) = z = (q, \mathcal{X}) \wedge \delta(q,\sigma)! \wedge (\exists w \in \Phi ^\pi (s)) \\
		& \sigma \in \mu(\xi(x_0, w) \cap Q_H) \\
		\Leftrightarrow 
		& \delta(q_0, s) =q \wedge \delta(q,\sigma)! \wedge (\exists w \in \Phi ^\pi (s)) \\
		& \sigma \in \mu(\xi(x_0, w) \cap Q_H) \\
		\Leftrightarrow 
		& \delta(q_0, s\sigma)! \wedge (\exists w \in \Phi ^\pi (s)) \sigma \in \mu(\xi(x_0, w) \cap Q_H) \\
		\Leftrightarrow 
		& s \sigma \in L(G)  \wedge  (\exists w \in \Phi ^\pi (s)) \sigma \in \mu(\xi(x_0, w) \cap Q_H ).
	\end{align*}
	\vspace{-5 mm}
	
	We further prove
	\vspace{-2 mm}
	\begin{equation} \label{Eq14}
		\begin{split}
			& s \sigma \in L_a({\cS^a_{CA}/G}) \\
			\Leftrightarrow 
			& s \in  L_a({\cS^a_{CA}/G})  \wedge s \sigma \in L(G) \wedge  (\exists w \in \Phi ^\pi (s)) \\
			& \sigma \in \cS_{CA} (w) \\
		\end{split}
	\end{equation}
	as follows.
		\vspace{-2 mm}
	\begin{align*}
		& s \sigma \in L_a({\cS^a_{CA}/G}) \\
		\Leftrightarrow 
		& s \in  L_a({\cS^a_{CA}/G}) \wedge s \sigma \in L(G) \wedge (\sigma \in \Sigma _{uc} \\
		& \vee (\exists w \in \Phi ^\pi (s))(\exists \gamma_a \in \cS ^a_{CA}(w)) \sigma \in \gamma_a) \\
		& (\mbox{by Equation  (\ref{Equation9})}) \\
		\Leftrightarrow 
		& s \in  L_a({\cS^a_{CA}/G}) \wedge s \sigma \in L(G) \wedge (\sigma \in \Sigma _{uc} \\
		& \vee (\exists w \in \Phi ^\pi (s))(\exists \gamma_a \in \Delta (\cS_{CA}(w))) \sigma \in \gamma_a) \\
		& (\mbox{by Equation  (\ref{Equation12})}) \\
		\Leftrightarrow 
		& s \in  L_a({\cS^a_{CA}/G}) \wedge s \sigma \in L(G) \wedge (\sigma \in \Sigma _{uc} \\
		& \vee (\exists w \in \Phi ^\pi (s))\sigma \in \cS_{CA}(w) \cup \Sigma ^a_c) \\
		& (\mbox{by Equation  (\ref{Equation11})}) \\
		\Leftrightarrow 
		& s \in  L_a({\cS^a_{CA}/G}) \wedge s \sigma \in L(G) \wedge  (\exists w \in \Phi ^\pi (s))\\
		& \sigma \in \cS_{CA}(w) \cup \Sigma _{uc} \cup \Sigma ^a_c \\
		\Leftrightarrow 
		& s \in  L_a({\cS^a_{CA}/G})  \wedge s \sigma \in L(G) \wedge  (\exists w \in \Phi ^\pi (s)) \\
		& \sigma \in \cS_{CA} (w) \\
		& (\mbox{by Equation  (\ref{Eq9})}) .
	\end{align*}
	
	Finally, we  prove that, for all $s \in \Sigma ^*$, 
	$$
	s \in L(H_{aug}) \Leftrightarrow s \in L_a({\cS^a_{CA}/G})
	$$
	by induction on the length $|s|$ of $s$.
	
	{\em Induction Base:}  Since the initial state $z_0 = (q_0, \{x_0\})$ is well defined, $\varepsilon \in  L(H_{aug})$. By the definition of  large language, $\varepsilon \in L_a({\cS^a_{CA}/G})$. 
	Therefore, for $|s| = 0$, that is, $s=\varepsilon$, we have
	$$
	s \in L(H_{aug}) \Leftrightarrow s \in L_a({\cS^a_{CA}/G}).
	$$
	
	{\em Induction Hypothesis:} Assume that for all $s \in \Sigma^*$,
	$|s| \leq m$,
	$$
	s \in L(H_{aug}) \Leftrightarrow s \in L_a({\cS^a_{CA}/G}).
	$$
	
	{\em Induction Step:} We show that for all $s \in \Sigma^*$, $\sigma \in \Sigma$, $|s\sigma| = m+1$,
	$$
	s \sigma \in L(H_{aug}) \Leftrightarrow s\sigma \in L_a({\cS^a_{CA}/G})
	$$
	as follows.
	\begin{align*}
		& s \sigma \in L_a({\cS^a_{CA}/G}) \\
		\Leftrightarrow 
		& s \in  L_a({\cS^a_{CA}/G})  \wedge s \sigma \in L(G) \wedge  (\exists w \in \Phi ^\pi (s)) \\
		& \sigma \in \cS_{CA} (w) \\
		& (\mbox{by Equation } (\ref{Eq14})) \\
		\Leftrightarrow 
		& s \in L(H_{aug}) \wedge s \sigma \in L(G) \wedge  (\exists w \in \Phi ^\pi (s)) \\
		& \sigma \in \cS_{CA} (w) \\
		&  (\mbox{by Induction Hypothesis}) \\
		\Leftrightarrow 
		& s \in L(H_{aug}) \wedge s \sigma \in L(G) \wedge  (\exists w \in \Phi ^\pi (s)) \\
		& \sigma \in \mu(SE^\pi_H(w)) \\
		& (\mbox{by Equation } (\ref{equL})) \\
		\Leftrightarrow 
		& s \in L(H_{aug}) \wedge s \sigma \in L(G) \wedge  (\exists w \in \Phi ^\pi (s)) \\
		& \sigma \in \mu(\xi (x_0, w) \cap Q_H) \\
		& (\mbox{by Proposition } \ref{PropositionSE})\\
		\Leftrightarrow 
		& s \in L(H_{aug}) \wedge s \sigma \in L(G) \wedge  (\exists w \in \Phi ^\pi (s)) \\
		& x= \xi (x_0, w) \wedge \sigma \in \mu( x\cap Q_H) \\
		\Leftrightarrow 
		& s \in  L(H_{aug})  \wedge s \sigma \in L(G) \\
		& \wedge(\exists x \in \mathcal{X}) \sigma \in \mu(x \cap Q_H) \\
		& (\mbox{by Equation } (\ref{rho3}))\\
		\Leftrightarrow 
		& s \in  L(H_{aug})  \wedge s \sigma \in L(G) \wedge q = \delta(q_0, s) \wedge \delta(q,\sigma)! \\
		& \wedge(\exists x \in \mathcal{X}) \sigma \in \mu(x \cap Q_H) \\
		\Leftrightarrow 
		& s \in  L(H_{aug})  \wedge s \sigma \in L(G) \wedge \rho(\rho(z_0, s),  \sigma)!\\
		& (\mbox{by Equation } (\ref{rho1}))\\
		\Leftrightarrow 
		& s \in  L(H_{aug})  \wedge s \sigma \in L(G) \wedge s \sigma \in L(H_{aug}) \\
		\Leftrightarrow 
		& s \sigma \in L(H_{aug})
	\end{align*}
\vspace{-10 mm}
\section{PROOF OF THEOREM \ref{theoaug4}}
\vspace{-2 mm}
	We first prove that $K$ is CA-observable (with respect to $L(G)$, $\Sigma _{o}$, $\Sigma ^a_o$ and $\Phi^\pi$) if and only if $L(H_{aug})$ $=$ $K$ as follows.
	\begin{align*}
		& K \mbox{ is CA-observable }\\
		\Leftrightarrow & L_a({\cS^a_{CA}/G}) = K\\
		&(\mbox{by Theorem } \ref{TheoremEC})\\
		\Leftrightarrow & L(H_{aug}) = K\\
		&(\mbox{by Theorem } \ref{theoaug})
	\end{align*}
	\vspace{-8 mm}
		
	It is not difficult to prove that $L(H_{aug}) = K$  if and only if
		\vspace{-3 mm}
	$$
	(\forall z=(q, \mathcal{X}) \in Z) \Gamma_{H_{aug}}(z) = \Gamma_H(q).
	$$
\vspace{-12 mm}
\section{PROOF OF THEOREM \ref{Proposition3}}
\vspace{-2 mm}
	We prove that CA-observability is not closed under intersection by the following counter example. Consider the same discrete event system $G$ (Fig. \ref{fig:G}), $\Sigma_o$, $\Sigma_o^a$, 
	and attack language $F_{tr}$ (Fig. \ref{fig:Ftr}) as in Example 1. Let $K_1$ be the $K$ in Fig. \ref{fig:K1}. It is shown in Example 1 that $K_1$ is CA-observable.
	
\begin{figure}[hbt!]
		\centering
		\includegraphics[scale=0.9]{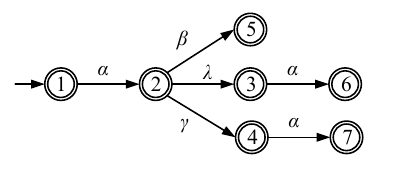}
		\caption{Specification language $K_2$ of the counter example in Theorem \ref{Proposition3}.  }
		\label{fig:K2}
	\end{figure}

	Let $K_2$ be the language generated by the automaton shown in Fig. \ref{fig:K2}. Similar to Example 1, it can be shown that $K_2$ is CA-observable (by exchanging $\beta$ and $\gamma$).
	
	\vspace{-5 mm}
		
\begin{figure}[hbt!]
		\centering
		\includegraphics[scale=0.9]{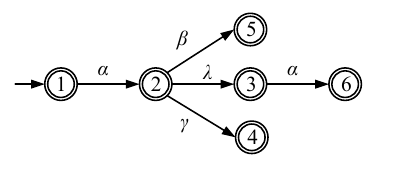}
		\caption{The intersection $K_1 \cap K_2$ of the counter example in Theorem \ref{Proposition3}.  }
		\label{fig:K12}
	\end{figure}

	The intersection $K_1 \cap K_2$ is generated by the automaton shown in Fig. \ref{fig:K12}.
	
	\vspace{-5 mm}
		
\begin{figure}[hbt!]
		\centering
		\includegraphics[scale=0.9]{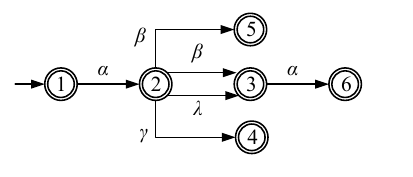}
		\caption{Nondeterministic automaton $H^\diamond$ for $K_1 \cap K_2$ of the counter example in Theorem \ref{Proposition3}.  }
		\label{fig:H12}
	\end{figure}

	The nondeterministic automaton $H^\diamond$ for $K_1 \cap K_2$ is shown in Fig. \ref{fig:H12}.
	
	\vspace{-2 mm}
		
\begin{figure}[hbt!]
		\centering
		\includegraphics[scale=0.9]{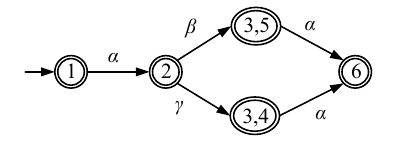}
		\caption{CA-observer $H_{obs}^\diamond$ for $K_1 \cap K_2$ of the counter example in Theorem \ref{Proposition3}.  }
		\label{fig:Ho12}
	\end{figure}

	The CA-observer $H_{obs}^\diamond$ for $K_1 \cap K_2$ is shown in Fig. \ref{fig:Ho12}. 
	
\begin{figure}[hbt!]
		\centering
		\includegraphics[scale=0.9]{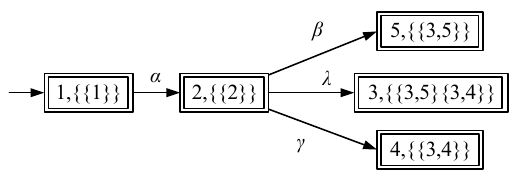}
		\caption{Augmented automaton  $H_{aug}$ for $K_1 \cap K_2$ of the counter example in Theorem \ref{Proposition3}.  }
		\label{fig:Ga12}
	\end{figure}

	The augmented automaton  $H_{aug}$ for $K_1 \cap K_2$ is shown in Fig. \ref{fig:Ga12}.
	
	Clearly, $L(H_{aug}) \not= K_1 \cap K_2$. Therefore, $K_1 \cap K_2$ is not CA-observable. This proves that CA-observability is not closed under intersection.
\vspace{-6 mm}
\section{PROOF OF THEOREM \ref{Proposition2}}
\begin{figure}[hbt!]
		\centering
		\includegraphics[scale=0.9]{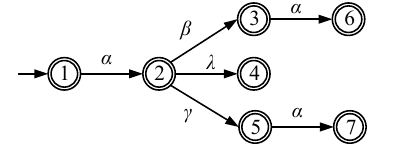}
		\caption{Discrete event system $G$ of the counter example in Theorem \ref{Proposition2}.  }
		\label{fig:GG}
	\end{figure}

\vspace{-2 mm}
	We prove that CA-observability is not closed under union by the following counter example. Consider the discrete event system $G$ shown in Fig. \ref{fig:GG}. 
	Assume that all events are observable, that is, $\Sigma = \Sigma_o = \{\alpha, \beta, \gamma, \lambda \}$. 
	
\begin{figure}[hbt!]
		\centering
		\includegraphics[scale=0.9]{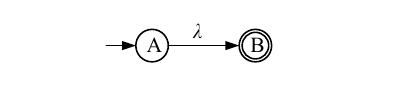}
		\caption{Attack language $F_{tr}$ for both $tr=(2, \beta, 3)$ and $tr=(2, \gamma, 5)$ of the counter example in Theorem \ref{Proposition2}.  }
		\label{fig:FFtr}
	\end{figure}

	Assume that $\beta$ and $\gamma$ may be attacked, that is, $\Sigma_o^a = \{\beta, \gamma \}$. 
	For both transition $tr=(2, \beta, 3) \in \delta^a$ and $tr=(2, \gamma, 5) \in \delta^a$, the attack language $F_{tr}$ is marked by the automaton shown in Fig. \ref{fig:FFtr}.
		\vspace{-3 mm}
		
\begin{figure}[hbt!]
		\centering
		\includegraphics[scale=0.9]{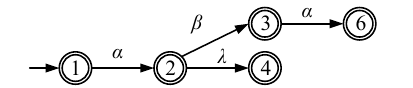}
		\caption{Specification language $K_1$ of  the counter example in Theorem \ref{Proposition2}.  }
		\label{fig:KK1}
	\end{figure}

	Let the specification language $K_1$ be generated by the automaton shown in Fig. \ref{fig:KK1}.
	
\begin{figure}[hbt!]
		\centering
		\includegraphics[scale=0.9]{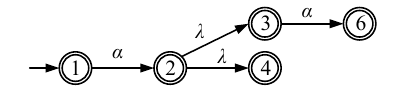}
		\caption{Nondeterministic automaton $H^\diamond$ for $K_1$ of the counter example in Theorem \ref{Proposition2}.  }
		\label{fig:HH1}
	\end{figure}

	The nondeterministic automaton $H^\diamond$ for $K_1$ is shown in Fig. \ref{fig:HH1}.
	
\begin{figure}[hbt!]
		\centering
		\includegraphics[scale=0.9]{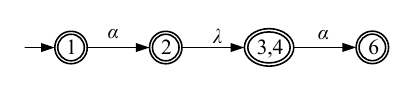}
		\caption{CA-observer $H_{obs}^\diamond$ for $K_1$ of the counter example in Theorem \ref{Proposition2}.  }
		\label{fig:HHo1}
	\end{figure}

	The CA-observer $H_{obs}^\diamond$ for $K_1$ is shown in Fig. \ref{fig:HHo1}.
	
\begin{figure}[hbt!]
		\centering
		\includegraphics[scale=0.9]{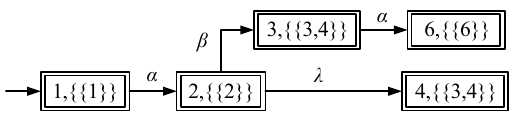}
		\caption{Augmented automaton  $H_{aug}$ for $K_1$ of the counter example in Theorem \ref{Proposition2}.  }
		\label{fig:GGa1}
	\end{figure}

	The augmented automaton  $H_{aug}$ for $K_1$ is shown in Fig. \ref{fig:GGa1}. Since, $L(H_{aug}) = K_1$, $K_1$ is  CA-observable. 
	
\begin{figure}[hbt!]
		\centering
		\includegraphics[scale=0.9]{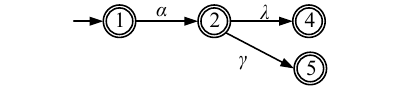}
		\caption{Specification language $K_2$ of  the counter example in Theorem \ref{Proposition2}.  }
		\label{fig:KK2}
	\end{figure}

	Let the specification language $K_2$ be generated by the automaton shown in Fig. \ref{fig:KK2}.
	
\begin{figure}[hbt!]
		\centering
		\includegraphics[scale=0.9]{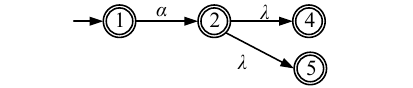}
		\caption{Nondeterministic automaton $H^\diamond$ for $K_2$ of the counter example in Theorem \ref{Proposition2}.  }
		\label{fig:HH2}
	\end{figure}

	The nondeterministic automaton $H^\diamond$ for $K_2$ is shown in Fig. \ref{fig:HH2}. 
	
\begin{figure}[hbt!]
		\centering
		\includegraphics[scale=0.9]{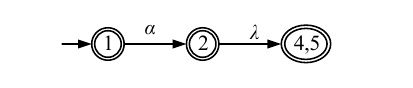}
		\caption{CA-observer $H_{obs}^\diamond$ for $K_2$ of the counter example in Theorem \ref{Proposition2}.  }
		\label{fig:HHo2}
	\end{figure}

	The CA-observer $H_{obs}^\diamond$ for $K_2$ is shown in Fig. \ref{fig:HHo2}. 
	
\begin{figure}[hbt!]
		\centering
		\includegraphics[scale=0.9]{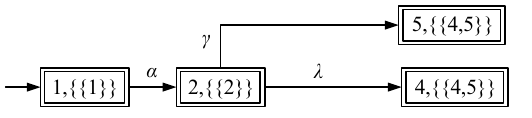}
		\caption{Augmented automaton  $H_{aug}$ for $K_2$ of the counter example in Theorem \ref{Proposition2}.  }
		\label{fig:GGa2}
	\end{figure}

	The augmented automaton $H_{aug}$ for $K_2$ is shown in Fig. \ref{fig:GGa2}. Since, $L(H_{aug}) = K_2$, $K_2$ is also CA-observable. 
	
\begin{figure}[hbt!]
		\centering
		\includegraphics[scale=0.9]{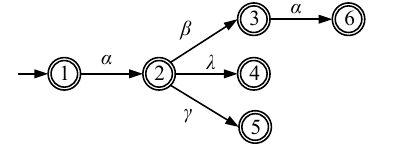}
		\caption{Specification language $K_1 \cup K_2$ of  the counter example in Theorem \ref{Proposition2}.  }
		\label{fig:KK12}
	\end{figure}

	Let us now consider the union of $K_1$ and $K_2$. The union $K_1 \cup K_2$ is generated by the automaton shown in Fig. \ref{fig:KK12}.
	
\begin{figure}[hbt!]
		\centering
		\includegraphics[scale=0.9]{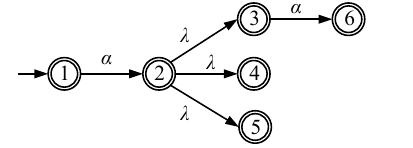}
		\caption{Nondeterministic automaton $H^\diamond$ for $K_1 \cup K_2$ of the counter example in Theorem \ref{Proposition2}.  }
		\label{fig:HH12}
	\end{figure}

	The nondeterministic automaton $H^\diamond$ for $K_1 \cup K_2$ is shown in Fig. \ref{fig:HH12}. 
	
\begin{figure}[hbt!]
		\centering
		\includegraphics[scale=0.9]{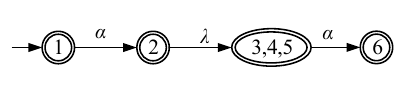}
		\caption{CA-observer $H_{obs}^\diamond$ for $K_1 \cup K_2$ of the counter example in Theorem \ref{Proposition2}.  }
		\label{fig:HHo12}
	\end{figure}

	The CA-observer $H_{obs}^\diamond$ for $K_1 \cup K_2$ is shown in Fig. \ref{fig:HHo12}. 
	
\begin{figure}[hbt!]
		\centering
		\includegraphics[scale=0.9]{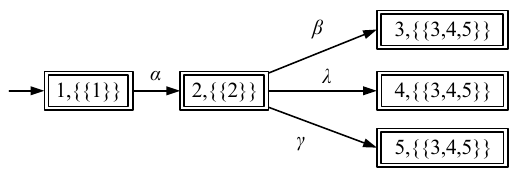}
		\caption{Augmented automaton  $H_{aug}$ for $K_1 \cup K_2$ of the counter example in Theorem \ref{Proposition2}.  }
		\label{fig:GGa12}
	\end{figure}

	The augmented automaton $H_{aug}$ for $K_1 \cup K_2$ is shown in Fig. \ref{fig:GGa12}. 
	
	Clearly, $L(H_{aug}) \not= K_1 \cup K_2$. Therefore, $K_1 \cup K_2$ is not CA-observable. This proves that CA-observability is not closed under union.

\section{PROOF OF THEOREM \ref{Theorem6}}
	Since $K^\uparrow$ is CA-controllable with respect to $L(G)$, $\Sigma _{uc}$, and $\Sigma ^a_c$, using a proof similar to that of Theorem \ref{theoaug}, it can be shown 
	$$
	L(H_{aug}^\uparrow) = L_a({\cS_{CA}^{\uparrow a}/G}).
	$$
	where CA-supervisor $\cS _{CA}^\uparrow : \Phi ^\pi ( L(G)) \rightarrow \Gamma$ is defined as follows. For $w \in \Phi ^\pi ( L(G))$,
	\begin{equation} \label{Equation13}
		\begin{split}
			SE^\pi_{H^\uparrow}(w)= & \{ q \in Q: (\exists s \in L(H^\uparrow)) \\
			& w \in \Phi ^\pi (s) \wedge \delta  (q_0, s)=q \} \\
			\cS _{CA}^\uparrow (w) = & \{\sigma \in \Sigma : (\forall q \in SE^\pi_{H^\uparrow} (w))q \in Q_H^\uparrow \\
			&\wedge \delta (q,\sigma) \in Q \Rightarrow \delta (q,\sigma) \in Q_H^\uparrow \} \\
			= & \mu ^\uparrow (SE^\pi_{H^\uparrow} (w)) .
		\end{split}
	\end{equation}
	
	By Theorem \ref{TheoremEC}, $L_a({\cS_{CA}^{\uparrow a}/G})$ is CA-controllable with respect to $L(G)$, $\Sigma _{uc}$, and $\Sigma ^a_c$ and CA-observable with respect to $L(G)$, $\Sigma _{o}$, $\Sigma ^a_o$ and $\Phi^\pi$. 
	Therefore, $L(H_{aug}^\uparrow)$ is a CA-controllable and CA-observable. 
	
	It remains to show that $L(H_{aug}^\uparrow)$ is a sublanguage of $K$, that is,
	$$
	L(H_{aug}^\uparrow) \subseteq K.
	$$
	We prove something stronger, that is, for all $s \in \Sigma ^*$, 
	$$
	s \in L(H_{aug}^\uparrow) \Rightarrow s \in K^\uparrow
	$$
	by induction on the length $|s|$ of $s$.
	
	{\em Induction Base:}  Since the initial state $z_0 = (q_0, \{x_0\})$ is well defined, $\varepsilon \in  L(H_{aug} ^\uparrow)$. Since $K$ is closed, $\varepsilon \in K$. 
	Therefore, for $|s| = 0$, that is, $s=\varepsilon$, we have
	$$
	s \in L(H_{aug}^\uparrow) \Rightarrow s \in K^\uparrow.
	$$
	
	{\em Induction Hypothesis:} Assume that for all $s \in \Sigma^*$,
	$|s| \leq m$,
	$$
	s \in L(H_{aug}^\uparrow) \Rightarrow s \in K^\uparrow.
	$$
	
	{\em Induction Step:} We show that for all $s \in \Sigma^*$, $\sigma \in \Sigma$, $|s\sigma| = m+1$,
	$$
	s \sigma \in L(H_{aug}^\uparrow) \Rightarrow s \sigma \in K^\uparrow
	$$
	as follows. Denote $z = \rho ^\uparrow (z_0, s) = (q, \mathcal{X})$,
	\begin{align*}
		& s \sigma \in L(H_{aug} ^\uparrow) \\
		\Leftrightarrow 
		& s \in  L(H_{aug} ^\uparrow)  \wedge s \sigma \in L(G) \wedge s \sigma \in L(H_{aug} ^\uparrow) \\
		\Leftrightarrow 
		& s \in  L(H_{aug} ^\uparrow)  \wedge s \sigma \in L(G) \wedge \rho ^\uparrow (z,  \sigma)! \\
		\Leftrightarrow 
		& s \in  L(H_{aug} ^\uparrow)  \wedge s \sigma \in L(G) \wedge q = \delta(q_0, s) \wedge \delta(q,\sigma)! \\
		& \wedge(\exists x \in \mathcal{X}) \sigma \in \mu ^\uparrow (x \cap Q_H) \\
		& (\mbox{by Equation (\ref{rho1}) for } H_{aug} ^\uparrow)\\
		\Leftrightarrow 
		& s \in L(H_{aug} ^\uparrow) \wedge s \sigma \in L(G) \wedge q = \delta(q_0, s) \wedge \delta(q,\sigma)! \\
		& \wedge (\exists w \in \Phi ^\pi (s)) x= \xi ^\uparrow (x_0, w) \wedge \sigma \in \mu ^\uparrow ( x\cap Q_H) \\
		& (\mbox{by Equation (\ref{rho3}) for } H_{aug} ^\uparrow)\\
		\Leftrightarrow 
		& s \in L(H_{aug} ^\uparrow) \wedge s \sigma \in L(G) \wedge q = \delta(q_0, s) \wedge \delta(q,\sigma)! \\
		& \wedge (\exists w \in \Phi ^\pi (s)) \sigma \in \mu ^\uparrow ( \xi ^\uparrow (x_0, w) \cap Q_H) \\
		\Leftrightarrow 
		& s \in L(H_{aug} ^\uparrow) \wedge s \sigma \in L(G) \wedge q = \delta(q_0, s) \wedge \delta(q,\sigma)! \\
		& \wedge (\exists w \in \Phi ^\pi (s)) \sigma \in \mu ^\uparrow ( SE^\pi_{H^\uparrow} (w) ) \\
		& (\mbox{by Proposition \ref{PropositionSE} for }  H_{aug} ^\uparrow) \\
		\Leftrightarrow 
		& s \in L(H_{aug} ^\uparrow) \wedge s \sigma \in L(G) \wedge q = \delta(q_0, s) \wedge \delta(q,\sigma)!  \\
		& \wedge  (\exists w \in \Phi ^\pi (s)) (\forall q' \in SE^\pi_{H^\uparrow} (w) )q' \in Q_H^\uparrow \\
		& \wedge \delta (q',\sigma) \in Q \Rightarrow \delta (q',\sigma) \in Q_H^\uparrow \\
		& (\mbox{by Equation } (\ref{Equation13})) \\
		\Rightarrow 
		& s \in L(H_{aug} ^\uparrow) \wedge s \sigma \in L(G) \wedge q = \delta(q_0, s) \wedge \delta(q,\sigma)!  \\
		&  \delta (q,\sigma) \in Q_H^\uparrow \\
		& (\mbox{because } (\forall w \in \Phi ^\pi (s)) q \in SE^\pi_{H^\uparrow} (w) \wedge q \in Q_H^\uparrow \\
		& \wedge \delta (q,\sigma) \in Q \mbox{ is true}) \\
		\Leftrightarrow 
		& s \in L(H_{aug} ^\uparrow) \wedge s \sigma \in L(G) \wedge q = \delta(q_0, s) \wedge \delta(q,\sigma)!  \\
		& \wedge \delta (\delta(q_0, s),\sigma) \in Q_H^\uparrow \\
		\Rightarrow 
		& s \in K^\uparrow \wedge s \sigma \in L(G) \wedge q = \delta(q_0, s) \wedge \delta(q,\sigma)!  \\
		& \wedge \delta (\delta(q_0, s),\sigma) \in Q_H^\uparrow \\
		& (\mbox{by Induction Hypothesis}) \\
		\Leftrightarrow 
		& s \in K^\uparrow \wedge s \sigma \in L(G) \wedge q = \delta(q_0, s) \wedge \delta(q,\sigma)!  \\
		& \wedge \delta (q_0, s \sigma) \in Q_H^\uparrow \\
		\Leftrightarrow 
		& s \in K^\uparrow \wedge s \sigma \in L(G) \wedge q = \delta(q_0, s) \wedge \delta(q,\sigma)!  \\
		& \wedge s \sigma \in K^\uparrow  \\
		\Leftrightarrow 
		& s \sigma \in K^\uparrow  .
	\end{align*}
\vspace{-10 mm}

\bibliographystyle{ieeetr}

\bibliography{DiagnosabilitySA}             
\vspace{-14 mm} 
\begin{IEEEbiography}[{\includegraphics[width=1in,height=1.25in,clip,keepaspectratio]{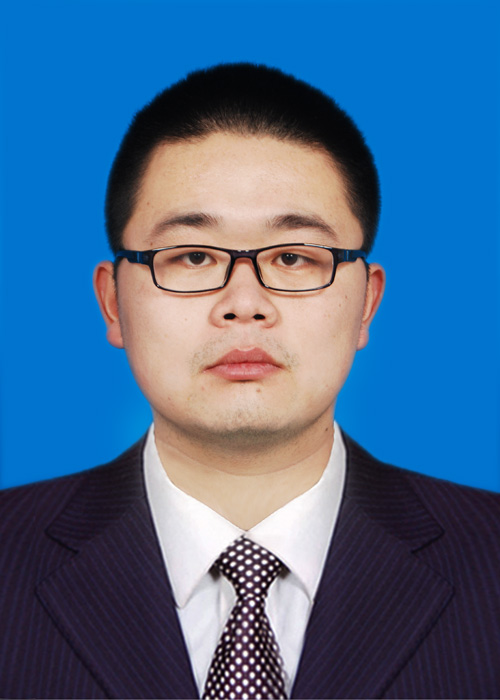}}]{Shengbao Zheng} 
	received  his Ph.D. degree in control theory and control engineering from Tongji University, Shanghai, China, in 2024 and the M.Eng. degree in control engineering from Northeastern University, Shenyang, China, in 2019 and the B.Eng. degree in automation from Shenyang University of Technology, Shenyang, China, in 2014. He has been with the School of Automation and Electrical Engineering , Shenyang Ligong University, Shenyang, Liaoning, China, where he is currently a lecturer. His current research interests include discrete event systems, cyber attacks and cyber-physical systems.
\end{IEEEbiography}
\vspace{-15 mm} 
\begin{IEEEbiography}[{\includegraphics[width=1in,height=1.25in,clip,keepaspectratio]{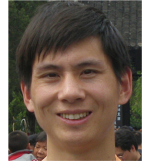}}]{Shaolong Shu} 
	(M’12–SM’15) received his B.Eng. degree in automatic control, and his Ph.D. degree in control theory and control engineering from Tongji University, Shanghai, China, in 2003 and 2008, respectively. Since July, 2008, he has been with the School of Electronics and Information Engineering, Tongji University, Shanghai, China, where he is currently a professor. From August, 2007 to February, 2008 and from April, 2014 to April, 2015, he was a visiting scholar in Wayne State University, Detroit, MI, USA. His main research interests include state estimation, fault-tolerant control and networked control of discrete event systems.	
\end{IEEEbiography}
\vspace{-13 mm} 
\begin{IEEEbiography}[{\includegraphics[width=1in,height=1.25in,clip,keepaspectratio]{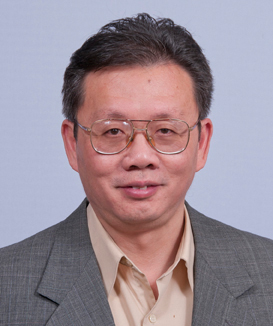}}]{Feng Lin} 
	(S'85-M'88-SM'07-F'09) received his B.Eng. degree in electrical engineering from Shanghai Jiao Tong University, Shanghai, China, in 1982, and the M.A.Sc. and Ph.D. degrees in electrical engineering from the University of Toronto, Toronto, ON, Canada, in 1984 and 1988, respectively. He was a Post-Doctoral Fellow with Harvard University, Cambridge, MA, USA, from 1987 to 1988. Since 1988, he has been with the Department of Electrical and Computer Engineering, Wayne State University, Detroit, MI, USA, where he is currently a Professor. His current research interests include discrete event systems, hybrid systems, neural networks, robust control, and their applications in alternative energy, biomedical systems, machine learning, and automotive control. He authored a book entitled ``Robust Control Design: An Optimal Control Approach'' and coauthored a paper that received a George Axelby outstanding paper award from the IEEE Control Systems Society. He was an associate editor of IEEE Transactions on Automatic Control.
	
\end{IEEEbiography}	
\end{document}